\documentclass[aps,twocolumn,floats,pre]{revtex4}
\usepackage{graphics,graphicx,epsfig}
\usepackage{amssymb,color}
\usepackage{epsf,epstopdf,wrapfig}

\begin{document}

\title{When are correlations strong?}

\author{Feraz Azhar$^{1,2}$ and William Bialek$^3$}
\affiliation{$^1$Department of Physics, University of California, Santa Barbara, California 93106\\
$^2$ Department of Neurosurgery, Brigham and Women's Hospital, Harvard Medical School, Boston, Massachusetts 02115\\
$^3$Joseph Henry Laboratories of Physics, Lewis--Sigler Institute for Integrative Genomics, and Princeton Center for Theoretical Science, Princeton University, Princeton, New Jersey 08544}

\date{\today}

\begin{abstract}
The inverse problem of statistical mechanics involves finding the minimal Hamiltonian that is consistent with some observed set of correlation functions.  This problem has received renewed interest in the analysis of biological networks; in particular, several such networks have been described successfully by maximum entropy models consistent with pairwise correlations.  These correlations are usually weak in an absolute sense (e.g., correlation coefficients $\sim 0.1$ or less), and this is sometimes taken as evidence against the existence of interesting collective behavior in the network.  If correlations are weak, it should be possible to capture their effects in perturbation theory, so we develop an expansion for the entropy of Ising systems in powers of the correlations, carrying this out to fourth order.  We then consider recent work on networks of neurons [Schneidman et al., {\em Nature} {\bf 440,} 1007 (2006); Tka\v{c}ik et al., arXiv:0912.5409 [q--bio.NC] (2009)], and show that even though all pairwise correlations are weak, the fact that these correlations are widespread means that their impact  on the network as a whole is not captured in the leading orders of perturbation theory.  
More positively, this means that recent successes of maximum entropy approaches are not simply the result of correlations being weak.
\end{abstract}

\maketitle

\section{Introduction}

Most of what is interesting about the phenomena of life results from interaction among large networks of elements---protein structures are stabilized by networks of interactions among amino acids, metabolism is governed by a network of enzymatic reactions, decisions about cell fate during embryonic development are determined by a network of genetic regulatory interactions, and our perceptions are shaped by dynamic interactions among networks of neurons.  Physicists have long hoped that the behavior of such large networks could be approached using ideas from statistical mechanics, and this idea has been explored most fully in the context of neural networks \cite{hopfield_82,amit_89}.  In contrast to the usual statistical mechanics problems, however, it is not clear how to measure the macroscopic ``thermodynamic'' properties of these networks.  On the other hand, a new generation of experiments is making it possible to observe something closer to the microscopic state of these networks, for example recording the activity of large numbers of neurons simultaneously \cite{segev+al_04,gunning+al_05,other}.  Given such data, what can we say about the global structure of the network?

Although experiments are continually improving, they will never get to the point that they can sample fully the state space of even modest sized networks.  What such data can provide, with high precision, is data on a finite set of correlation functions or expectation values, or the distributions of some small set of order parameters.  To make progress toward a global description of the network, we need to solve an inverse problem.  In the language of statistical mechanics, we are given the expectation values of various operators, and we need to infer the underlying Hamiltonian.  In general, of course, this is ill--posed.  Recently, a number of groups have explored the possibility that this inverse problem can be successfully regularized using the classical idea of maximum entropy \cite{jaynes_57}.  

To make these ideas concrete, note that the electrical activity in networks of neurons consists of discrete, identical pulses termed action potentials or spikes \cite{adrian_26,adrian+zotterman_26a,spikes}.  In a small window of time, each neuron either generates one spike or remains silent, so that the state of the system is described naturally by Ising spins, spin up for a spike and spin down for silence.  Knowing the mean rate at which each cell generates spikes is equivalent to knowing the mean magnetization of each spin, and the probability of two cells generating spikes in the same small window of time is related to the spin--spin correlation function.  The maximum entropy model consistent with knowledge of the mean spike rates and pairwise correlations is then an Ising model with pairwise interactions \cite{schneidman+al_03,schneidman+al_06}, and the relevant inverse problem is to determine the magnetic fields and spin--spin interactions from measurements of the magnetization and two--point correlation functions.  In general these systems are inhomogeneous and likely to be glassy, since neurons can be both positively and negatively correlated.  

Interest in the inverse statistical mechanics approach to biological networks has been raised by several demonstrations that maximum entropy models built from pairwise correlations succeed in capturing the higher--order structure of these systems.  Examples include neurons in the retina \cite{schneidman+al_06,tkacik+al_06,shlens+al_06,tkacik+al_09}, in cultured networks \cite{schneidman+al_06,tang+al_08}, and in the cortex \cite{yu+al_08,other_neurons_2}.  An independent stream of work has shown that functional proteins can be constructed by drawing randomly from an ensemble of amino acid sequences that reproduces the correlations between substitutions at pairs of sites across known families of proteins \cite{lockless+ranganathan_99,suel+al_03,socolich+al_05,russ+al_05}.  This construction is, in certain limits, equivalent to a maximum entropy model \cite{bialek+ranganathan_07}, and this approach has now been applied to wide range of different proteins \cite{weigt+al_09,mora+al_09,halabi+al_09}.  Pairwise maximum entropy models have also been used to describe biochemical \cite{tkacik_07} and genetic \cite{psu_group} networks, and even the spelling rules for four letter words \cite{stephens+bialek_08}.

The full power of statistical mechanics approaches lies in the limit of large networks.  Generations of theoretical studies have led us to hope for interesting collective behavior in these systems, which of course becomes clear only in the large $N$ limit.   The maximum entropy construction provides a bridge between these theoretical expectations and real data \cite{schneidman+al_06,tkacik+al_06}.  The search for collective effects in large networks is also a subject of controversy, and settling these controversies will require actually solving the inverse problem in large systems.  Recent work suggests some promising approaches, but also highlights the difficulties of the problem \cite{broderick+al_07,mezard+mora_08,sessak+monasson_08}.

Here we try to make progress on something more modest than the full inverse problem.  We start with the observation that individual pairwise correlations often are weak in an absolute sense; for example, the correlation coefficient between the activity of two neurons typically is $C\sim 0.1$ or less.  This apparently weak correlation suggests that the effects of correlations will be small, and indeed if one looks at small groups of neurons this must be true. More formally, if we believe that correlations are weak, then it should be possible to capture the impact of these correlations in perturbation theory.  Here we develop this perturbation theory, evaluating the entropy of an Ising system out to fourth order in the spin--spin correlations \cite{note}.  We apply our results to re--analyze the correlations among neurons in the vertebrate retina \cite{schneidman+al_06}.

Our primary conclusions are negative:   real networks of neurons in the retina are outside the regime in which we can expect the leading orders of perturbation theory to capture the impact of the measured correlations, and we argue that this is true more generally for biological networks.    But even this negative result is important, because it shows us that the successes of maximum entropy models thus far are not simply the result of correlations being weak, so that even in groups of 20 of 40 neurons we are seeing meaningful hints of the emergent, collective behavior predicted by these models.  The perturbation expansion also highlights the difficulties of defining a thermodynamic limit for these systems.

\section{Calculation of the Entropy\label{generalcalculation}}

We are interested in a system of $N$ spins $\{\sigma_{\rm i}\}$, which represent the states of a biological network.  As noted above, this representation is especially simple for neurons, where $\sigma_{\rm i}=+1$ marks the occurrence of an action potential from neuron $\rm i$ in a small time window, while $\sigma_{\rm i} = -1$ indicates that neuron $\rm i$ is silent.  Although the problem is quite general, when we want to speak concretely we will use such networks of neurons as our example.

Once the number of elements $N$ in our system is large, no reasonable experiment can lead directly to an estimate of the full probability distribution $P(\{\sigma_{\rm i}\})$ describing the states of the system as a whole.  What we can hope to measure are expectation values for low--order operators, such as the mean magnetization of each spin,
\begin{equation}\label{onepoint}
\langle\sigma_{\rm i}\rangle = \sum_{\{\sigma_{\rm i}\}}P(\{\sigma_{\rm i}\})\sigma_{\rm i} ,
\end{equation}
and the spin--spin correlation,
\begin{equation}\label{twopoint}
\langle\sigma_{\rm i}\sigma_{\rm j}\rangle = \sum_{\{\sigma_{\rm i} \}}P(\{\sigma_{\rm i}\})\sigma_{\rm i} \sigma_{\rm j}  .
\end{equation}

In the maximum entropy formulation, one constructs a distribution which maximizes the entropy $S[P(\{\sigma_{\rm i}\})]$, subject to constraints. The entropy is defined as usual by %
\begin{equation}\label{shannonentropy}
S[P(\{\sigma_{\rm i}\})] \equiv  -\sum_{\{\sigma_{\rm i}\}}P(\{\sigma_{\rm i}\})\ln \left[P(\{\sigma_{\rm i}\})\right] ,
\end{equation}
where we measure entropy in nats unless stated otherwise.
One can in principle write down a maximum entropy distribution consistent with higher order correlations, but if we keep just the one--point and two--point expectation values then the
solution to the resulting constrained maximization problem is
\begin{equation}\label{ising}
P(\{\sigma_{\rm i}\})=\frac{1}{Z(\{h_{\rm i}, J_{\rm ij}\})}\exp\left(\sum_{\rm i} h_{\rm i}\sigma_{\rm i}+\frac{1}{2}\sum_{\rm i\neq j}J_{\rm ij}\sigma_{\rm i}\sigma_{\rm j}\right)
\end{equation}
where $Z(\{h_{\rm i}, J_{\rm ij}\})$, the partition function, is given by
\begin{equation}\label{isingpartition}
Z(\{h_{\rm i}, J_{\rm ij}\}) = \sum_{\{\sigma_{\rm i}\}}\exp\left(\sum_{\rm i} h_{\rm i}\sigma_{\rm i}+\frac{1}{2}\sum_{\rm i\neq j}J_{\rm ij}\sigma_{\rm i}\sigma_{\rm j}\right).
\end{equation}
The numbers $\{h_{\rm i}, J_{\rm ij}\}$ are Lagrange multipliers which are fixed by imposing the constraints in Eq's (\ref{onepoint}) and (\ref{twopoint}). One immediately recognizes Eq (\ref{ising}) as the Ising model where the interactions $\{J_{\rm ij}\}$ exist (potentially) between all pairs of spins.

The difficulty in computing the entropy $S[P(\{\sigma_{\rm i}\})]$ directly from the distribution in Eq (\ref{ising}) is that of imposing the constraints in Eq's (\ref{onepoint}) and (\ref{twopoint}).   This corresponds to solving the $N(N+1)/2$ simultaneous equations
\begin{eqnarray}
\langle \sigma_{\rm i} \rangle &=& {{\partial \ln Z(\{h_{\rm i}, J_{\rm ij}\}) }\over{\partial h_{\rm i}}}\\
\langle \sigma_{\rm i} \sigma_{\rm j}  \rangle &=& {{\partial \ln Z(\{h_{\rm i}, J_{\rm ij}\}) }\over{\partial J_{\rm ij}}} .
\end{eqnarray}
Our goal is to develop a perturbative approach to this problem.    
At the risk of being pedantic, we present the development in some detail, hopefully making the discussion accessible to a broader audience with interests in biological networks.

\subsection{The General Case}
\label{generic}

Let's start with a very general approach, in which we imagine that we know the expectation values for some set of operators $\hat O_\mu (\{\sigma_{\rm i}\})$, $\mu = 1, 2, \cdots , K$.  Then the partition function for the maximum entropy distribution takes the form

\begin{equation}\label{generalpartition}
Z(\{g_{\mu}\})=\sum_{\{\sigma_{\rm i}\}}\exp\left[\sum_{\mu =1}^K g_{\mu}\hat{O}_{\mu}(\{\sigma_{\rm i}\})\right] .
\end{equation}
The $\{g_{\mu}\}$ represent the coupling constants of the system, which arise as Lagrange multipliers in the constrained maximization problem from which this partition function originated.  The coupling constants are determined by the $K$ simultaneous equations
\begin{equation}
\langle \hat{O}_{\mu}(\{\sigma_{\rm i}\})\rangle = {{\partial\ln Z}\over{\partial g_\mu}} .
\end{equation}
We assume that there is some `zero order' condition in which the expectation values are $\langle \hat{O}_{\mu}(\{\sigma_{\rm i}\})\rangle^{(0)}$ and the corresponding coupling constants are $g_\mu^0$.  If we observe that expectation values are slightly different from their zero order values, this should have a proportionally small effect on the entropy, and this is what we want to calculate in perturbation theory.

As is usual in statistical mechanics, we can relate the entropy to derivatives of the free energy,
\begin{equation}
S(\{g_{\mu}\})=\ln Z(\{g_{\mu}\}) -\sum_\mu g_{\mu}\frac{\partial\ln Z(\{g_{\mu}\})}{\partial g_{\mu}} .
\end{equation}
Note that, in this view, entropy is a function of the coupling constants, and only implicitly a function of the measured expectation values.
To make the dependence on expectation values explicit, we consider
\begin{eqnarray}
{{\partial S}\over{\partial \langle \hat O_\mu\rangle}} &=&
\sum_\nu {{\partial S}\over{\partial g_\nu}} {{\partial g_\nu}\over
{\partial \langle \hat O_\mu\rangle}}\\
&=& \sum_\nu \left[ - \sum_\lambda g_\lambda {{\partial  \langle \hat O_\lambda\rangle}
\over{\partial g_\nu }}
\right] {{\partial g_\nu}\over
{\partial \langle \hat O_\mu\rangle}}\\
&=& - \sum_\lambda g_\lambda \left[\sum_\nu {{\partial  \langle \hat O_\lambda\rangle}
\over{\partial g_\nu }}{{\partial g_\nu}\over
{\partial \langle \hat O_\mu\rangle}}
\right] \\
&=& - g_\mu . \label{entropycoupling}
\end{eqnarray}
To use this expression we should view the coupling constants as functions of the expectation values, $g_\mu = g_\mu ( \{ \langle O_\lambda \} )$.  In the zero order state we have
\begin{equation}
g_\mu = g_\mu^0 \equiv g_\mu ( \{ \langle O_\lambda \rangle^{(0)} \} ), 
\end{equation}
and we measure the deviations from this state as
\begin{equation}\label{couplingsdeviation}
\delta g_{\mu} = g_{\mu} - g_{\mu}^{0} .
\end{equation}
Similarly, we define
the deviation of the operators from their zeroth order values to be
\begin{equation}\label{deviationofbig}
\Delta\hat{O}_{\mu}\equiv \hat{O}_{\mu}-\langle\hat{O}_{\mu}\rangle ^{(0)} .
\end{equation}
Then the entropy in the state we are interested can be found by integrating, starting from the zero order state:
\begin{eqnarray}
S &=& S(\{\langle\hat{O}_{\alpha}\rangle ^{(0)}\})-\sum_{\mu}g_{\mu}^{0}\langle\Delta\hat{O}_{\mu}\rangle
\nonumber\\
 & & -\sum_{\alpha}\int_{\{0\}}^{\{\langle\Delta\hat{O}_{\mu}\rangle\}}d \langle\Delta\hat{O}_{\alpha}\rangle\delta g_{\alpha} ,
\label{S-intg}
\end{eqnarray}
where $S(\{\langle\hat{O}_{\alpha}\rangle ^{(0)}\})$ is the entropy of the zeroth order distribution.

Now our task is clear---we need to develop a perturbation theory for the coupling constants themselves.  The zeroth order couplings define expectation values with respect to the induced zeroth order distribution in the usual way,
\begin{eqnarray}
\Big\langle\cdots\Big\rangle ^{(0)} &\equiv& \frac{1}{Z_{0}(\{g_{\mu}^{0}\})}\sum_{\{\sigma_{\rm i}\}}\exp \left[\sum_{\mu}g_{\mu}^{0}\;\hat{O}_{\mu}(\{\sigma_{\rm i}\})\right]\Big(\cdots\Big)\nonumber\\
&&
\label{operatorsdeviation}
\\
Z_{0}(\{g_{\mu}^{0}\}) &\equiv& \sum_{\{\sigma_{\rm i}\}}\exp \left[\sum_{\mu}g_{\mu}^{0}\;\hat{O}_{\mu}(\{\sigma_{\rm i}\})\right].
\end{eqnarray}
Using these definitions, one can rewrite the partition function as
\begin{eqnarray}
Z(\{g_{\mu}\})&=&Z_{0}(\{g_{\mu}^{0}\})\exp\left(\sum_{\mu}\delta g_{\mu}\langle\hat{O}_{\mu}\rangle ^{(0)}\right)
\nonumber\\
&&\,\,\,\,\,\,\,\,\times\Bigg\langle\exp \left[\sum_{\mu}\delta g_{\mu}\;\Delta\hat{O}_{\mu}\right]\Bigg\rangle ^{(0)}
\label{partitionbeforecumulant}
\end{eqnarray}
where for convenience we drop the explicit dependence of our operators on the binary variables.   From this expression we use the cumulant expansion to develop $\ln Z$ as a power series in $\delta g$, and then differentiate to obtain the expectation values.  The result is that
\begin{widetext}
\begin{equation}
\langle \hat O_\mu\rangle = \langle \hat O_\mu\rangle^{(0)}
+ \langle \Delta \hat O_\mu \Delta \hat O_\nu \rangle_c^{(0)}
 \delta g_\nu
+ {1\over {2!}} \langle \Delta \hat O_\mu \Delta \hat O_\nu \Delta \hat O_\lambda \rangle_c^{(0)} \delta g_\nu\delta g_\lambda
+{1\over {3!}}
\langle \Delta \hat O_\mu \Delta \hat O_\nu \Delta \hat O_\lambda \Delta \hat O_\rho\rangle_c^{(0)} \delta g_\nu\delta g_\lambda\delta g_\rho
+ \cdots , 
\end{equation}
where we sum over repeated indices for the remainder of this section. We can rewrite this as
\begin{equation}
\delta g_{\alpha}
=
(\chi ^{-1})_{\alpha\mu}\langle\Delta\hat{O}_{\mu}\rangle
-\frac{1}{2!}(\chi ^{-1})_{\alpha\mu}\delta g_{\nu}\delta g_{\lambda}\langle\Delta\hat{O}_{\mu}\Delta\hat{O}_{\nu}\Delta\hat{O}_{\lambda}\rangle_{c}^{(0)}
-\frac{1}{3!}(\chi ^{-1})_{\alpha\mu}\delta g_{\nu}\delta g_{\lambda}\delta g_{\rho}\langle\Delta\hat{O}_{\mu}\Delta\hat{O}_{\nu}\Delta\hat{O}_{\lambda}\Delta\hat{O}_{\rho}\rangle_{c}^{(0)}+ \cdots ,
\label{yoyo}
\end{equation}
where we identify the susceptibility
\begin{equation}\label{susceptibility}
\chi_{\mu\nu}
\equiv
\langle\Delta\hat{O}_{\mu}\Delta\hat{O}_{\nu}\rangle_{c}^{(0)}.
\end{equation}
Iterating this series expansion, we find
\begin{eqnarray}
\delta g_{\alpha}&=&(\chi ^{-1})_{\alpha\mu}\langle\Delta\hat{O}_{\mu}\rangle 
- \frac{1}{2}(\chi ^{-1})_{\alpha\mu}(\chi ^{-1})_{\beta\nu}(\chi ^{-1})_{\gamma\lambda}\langle\Delta\hat{O}_{\nu}\rangle\langle\Delta\hat{O}_{\lambda}\rangle\langle\Delta\hat{O}_{\mu}\Delta\hat{O}_{\beta}\Delta\hat{O}_{\gamma}\rangle_{c}^{(0)}
\nonumber \\& &\,\,\,\,\,\,\,\,\,\,
- \frac{1}{3!}(\chi ^{-1})_{\alpha\mu}(\chi ^{-1})_{\beta\nu}(\chi ^{-1})_{\gamma\lambda}(\chi ^{-1})_{\delta\rho}\langle\Delta\hat{O}_{\nu}\rangle\langle\Delta\hat{O}_{\lambda}\rangle\langle\Delta\hat{O}_{\rho}\rangle
\langle\Delta\hat{O}_{\mu}\Delta\hat{O}_{\beta}\Delta\hat{O}_{\gamma}\Delta\hat{O}_{\delta}\rangle_{c}^{(0)}
\nonumber \\& &\,\,\,\,\,\,\,\,\,\,
+ \frac{1}{2}(\chi ^{-1})_{\alpha\mu}(\chi ^{-1})_{\beta\nu}(\chi ^{-1})_{\gamma\lambda}(\chi ^{-1})_{\delta\rho}(\chi ^{-1})_{\epsilon\sigma}\langle\Delta\hat{O}_{\nu}\rangle\langle\Delta\hat{O}_{\rho}\rangle\langle\Delta\hat{O}_{\sigma}\rangle
\langle\Delta\hat{O}_{\mu}\Delta\hat{O}_{\beta}\Delta\hat{O}_{\gamma}\rangle_{c}^{(0)}\langle\Delta\hat{O}_{\lambda}\Delta\hat{O}_{\delta}\Delta\hat{O}_{\epsilon}\rangle_{c}^{(0)}
\nonumber\\
&&\,\,\,\,\,\,\,\,\, +\cdots.
\label{finaldeltag}
\end{eqnarray}

Having obtained an expression for the couplings perturbatively, we use the coupling constant integration [Eq (\ref{S-intg})] to generate an expansion for the entropy.  The result is
\begin{eqnarray}
S &=& S(\{\langle\hat{O}_{\alpha}\rangle ^{0}\})-g_{\mu}^{0}\langle\Delta\hat{O}_{\mu}\rangle -\frac{1}{2}(\chi ^{-1})_{\alpha\mu}\langle\Delta\hat{O}_{\alpha}\rangle\langle\Delta\hat{O}_{\mu}\rangle\nonumber\\
& &+\frac{1}{6}(\chi ^{-1})_{\alpha\mu}(\chi ^{-1})_{\beta\nu}(\chi ^{-1})_{\gamma\lambda}\langle\Delta\hat{O}_{\alpha}\rangle\langle\Delta\hat{O}_{\nu}\rangle\langle\Delta\hat{O}_{\lambda}\rangle\langle\Delta\hat{O}_{\mu}\Delta\hat{O}_{\beta}\Delta\hat{O}_{\gamma}\rangle_{c}^{(0)}\nonumber\\
& &-\frac{1}{8} (\chi ^{-1})_{\alpha\mu}(\chi ^{-1})_{\beta\nu}(\chi ^{-1})_{\gamma\lambda}(\chi ^{-1})_{\delta\rho}(\chi ^{-1})_{\epsilon\sigma}\langle\Delta\hat{O}_{\alpha}\rangle\langle\Delta\hat{O}_{\nu}\rangle\langle\Delta\hat{O}_{\rho}\rangle\langle\Delta\hat{O}_{\sigma}\rangle \langle\Delta\hat{O}_{\mu}\Delta\hat{O}_{\beta}\Delta\hat{O}_{\gamma}\rangle_{c}^{(0)}\langle\Delta\hat{O}_{\lambda}\Delta\hat{O}_{\delta}\Delta\hat{O}_{\epsilon}\rangle_{c}^{(0)}
\nonumber\\
& &+\frac{1}{24}(\chi ^{-1})_{\alpha\mu}(\chi ^{-1})_{\beta\nu}(\chi ^{-1})_{\gamma\lambda}(\chi ^{-1})_{\delta\rho}\langle\Delta\hat{O}_{\alpha}\rangle\langle\Delta\hat{O}_{\nu}\rangle\langle\Delta\hat{O}_{\lambda}\rangle\langle\Delta\hat{O}_{\rho}\rangle\langle\Delta\hat{O}_{\mu}\Delta\hat{O}_{\beta}\Delta\hat{O}_{\gamma}\Delta\hat{O}_{\delta}\rangle_{c}^{(0)}\nonumber\\
&&\,\,\,\,\,\,\,\,\,+\cdots .
\label{finalgeneralentropy}
\end{eqnarray}
\end{widetext}
These results allow us to express the entropy---or, more precisely, the maximum possible entropy---as a function of experimentally observable expectation values, assuming that these are close to some reference state which we understand exactly.

\subsection{The Pairwise Maximum Entropy Model}

In the pairwise maximum entropy model one assumes that the operators $\hat{O}_{\mu}$ take on two distinct expressions depending on their index. In the first sector, $\mu = {\rm i} \Rightarrow \hat{O}_{\mu}=\sigma_{\rm i}$ and in the second $\mu = {\rm ij} \Rightarrow \hat{O}_{\mu}=\sigma_{\rm i}\sigma_{\rm j}$. The general partition function of the last section [Eq (\ref{generalpartition})] then reduces to the partition function of the Ising model in Eq (\ref{isingpartition}). One can write down the entropy for this latter partition function in perturbation theory in terms of empirical quantities, namely, in terms of the one- and two-point correlation functions $\{\langle\sigma_{\rm i}\rangle\}$ and $\{\langle\sigma_{\rm i}\sigma_{\rm j}\rangle\}$ respectively. The final form of this entropy is that of our earlier result, Eq (\ref{finalgeneralentropy}). Here, we rewrite Eq (\ref{finalgeneralentropy}) in terms of quantities defined by the pairwise maximum entropy model, leaving the details of the calculation to the appendix.

We begin with a form for the partition function which re-expresses the operators appearing in the pairwise construction such that their expectation values vanish in  the zeroth order distribution. Namely we consider the following variant of Eq (\ref{isingpartition}),
\begin{equation}\label{isingpartition2}
Z(\{h_{\rm i}, J_{\rm ij}\}) = \sum_{\{\sigma_{\rm i}\}}\exp\left(\sum_{\rm i} h_{\rm i}\delta\sigma_{\rm i}+\sum_{\rm i < j}J_{\rm ij}\delta\sigma_{\rm i}\delta\sigma_{\rm j}\right).
\end{equation}
where $\delta\sigma_{\rm i} = \sigma_{\rm i} - \langle \sigma_{\rm i}\rangle^{(0)}$.  The zero order coupling constants $\{g_{\mu}^{0}\}$  correspond to a noninteracting model, $J_{\rm ij} = 0$, with the $\{h_{\rm i}^{0}\}$ chosen to reproduce the observed mean magnetizations,
\begin{equation}
\langle\sigma_{\rm i}\rangle = \langle\sigma_{\rm i}\rangle^{(0)} =\tanh(h_{\rm i}^{0}).
\end{equation}

One can rewrite the partition function in Eq (\ref{isingpartition2}) as 
\begin{widetext}
\begin{equation}\label{isingdelta}
Z(\{h_{\rm i}, J_{\rm ij}\})=Z_{0}(\{h_{\rm i}^{0}\})\exp \left(-\sum_{\rm i}h_{\rm i}^{0}\langle\sigma_{\rm i}\rangle ^{(0)}\right)\Bigg\langle\exp\left(\sum_{\rm i}\delta h_{\rm i}\delta\sigma_{\rm i}+\sum_{\rm i < j}J_{\rm ij}\delta\sigma_{\rm i}\delta\sigma_{\rm j}\right)\Bigg\rangle ^{(0)},
\end{equation}
\end{widetext}
where 
\begin{equation}
Z_{0}(\{h_{\rm i}^{0}\})
\equiv
\sum_{\{\sigma_{\rm i}\}}\exp \left(\sum_{\rm i}h_{\rm i}^{0}\sigma_{\rm i}\right) .
\end{equation}
As usual, zero order expectation values are defined by
\begin{equation}
\Big\langle\cdots\Big\rangle ^{(0)}
\equiv
\frac{1}{Z_{0}(\{h_{\rm i}^{0}\})}\sum_{\{\sigma_{\rm i}\}}\exp \left(\sum_{\rm i}h_{\rm i}^{0}\sigma_{\rm i}\right)\Big(\cdots\Big).
\end{equation}

Proceeding as in the general case, we obtain the analogue of Eq (\ref{finalgeneralentropy}). The result can depend only on the experimental one point correlations $\{\langle\sigma_{\rm i}\rangle^{(0)} = \langle\sigma_{\rm i}\rangle\}$ and the experimental two point correlations, which we summarize by the correlation coefficients
\begin{equation}\label{correlations}
C_{\rm ij}
\equiv
\frac{\langle\delta\sigma_{\rm i}\delta\sigma_{\rm j}\rangle}{\sqrt{\langle\left(\delta\sigma_{\rm i}\right)^{2}\rangle ^{(0)}}\sqrt{\langle\left(\delta\sigma_{\rm j}\right)^{2}\rangle ^{(0)}}}.
\end{equation}
We write the entropy as  
\begin{equation}
S(\{\langle\sigma_{\rm i}\rangle^{(0)}, C_{\rm ij}\}) = S_{0}(\{\langle\sigma_{\rm i}\rangle^{(0)}\})+\Delta S(\{\langle\sigma_{\rm i}\rangle^{(0)}, C_{\rm ij}\}) ,
\end{equation}
where 
\begin{eqnarray}
S_0 &=& N + \frac{1}{\ln {2}}\sum_{\rm i=1}^{N}\ln\left(\cosh\left(\tanh ^{-1}(\langle\sigma_{\rm i}\rangle ^{(0)})\right)\right)\nonumber\\
&&\,\,\,\,\,-\frac{1}{\ln 2}\sum_{\rm i=1}^{N}\langle\sigma_{\rm i}\rangle ^{(0)}\tanh ^{-1}(\langle\sigma_{\rm i}\rangle ^{(0)}) \;\;\textrm{bits} 
\end{eqnarray}
 is the entropy of the noninteracting system, and then collect terms with successive powers of the $C_{\rm ij}$:
\begin{equation}
\Delta S =  \Delta S_{2}+\Delta S_{3}+\cdots ,
\end{equation}
where 
\begin{widetext}
\begin{eqnarray}
\Delta S_{2}(\{C_{\rm ij}\}) &=   & -\frac{1}{4 \ln 2}\sum_{\rm i\neq j}{C_{\rm ij}}^{2} \;\;\textrm{bits}\\
\Delta S_{3}(\{\langle\sigma_{\rm i}\rangle^{(0)}, C_{\rm ij}\}) &=   & + \frac{1}{3!\ln 2}\sum_{\rm i\neq j\; j\neq l\; i\neq l}C_{\rm ij}C_{\rm jl}C_{\rm li}  + \frac{1}{3\ln 2}\sum_{\rm i\neq j}{C_{\rm ij}}^3\left[\frac{\langle\sigma_{\rm i}\rangle ^{(0)}}{\left(\delta\sigma_{\rm i}\right)^{(0)}_{\rm rms}}\right]\left[\frac{\langle\sigma_{\rm j}\rangle ^{(0)}}{\left(\delta\sigma_{\rm j}\right)^{(0)}_{\rm rms}}\right]  \;\;\textrm{bits} .
\end{eqnarray}
\end{widetext}
Details of the computation, including the fourth order term $\Delta S_4$, are collected in the Appendices.

\subsection{Remarks on the thermodynamic limit}

Many biological networks are large, and it is tempting to think that the essence of their behavior can be derived in the thermodynamic limit, $N\rightarrow \infty$.  We expect that, in this limit, the entropy is extensive, that is $S \propto N$.  But we have to be careful about how we define this limit, and what is held fixed as $N$ varies.

To illustrate the problem, consider the entropy to second order in the correlations (in bits),
\begin{equation}
S \approx S_0 - {1\over{4\ln 2}}\sum_{\rm i\neq j} {C_{\rm ij}}^{2} .
\end{equation}
As $N$ becomes large, we can write this as
\begin{equation}
S \approx S_0 - {{N(N-1)}\over{4\ln 2}} \langle C^2\rangle ,
\label{avgC2}
\end{equation}
where $\langle C^2\rangle$ denotes the average squared correlation coefficient in the network.  It is the entropy per spin which should be finite as $N\rightarrow \infty$, 
\begin{equation}
{S\over N} \rightarrow {{S_0}\over N} - {{N\langle C^2\rangle}\over{4\ln 2}} .
\end{equation}
To enforce the existence of a thermodynamic limit, it is tempting to say that we must have $N\langle C^2\rangle$ be finite as $N$ grows large.
The difficulty is that $\langle C^2\rangle$ is an experimental quantity, not something we are free to adjust theoretically.

We recall that, if we are studying a large system with a well defined geometry and a finite correlation length, then we expect correlations to decay with the distance between spins.  Roughly speaking, in dimension $d$ we expect that of the $\sim N^2$ pairs of spins that we could choose, only $\sim N (\xi/a)^d$ have significant correlations, where $a$ is the lattice spacing or typical distance between spins.  Thus the mean square correlation will scale as 
\begin{equation}
\langle C^2 \rangle \sim C_0^2 {{N (\xi/a)^d}\over{N^2}} \sim {{C_0^2 (\xi/a)^d}\over N} ,
\end{equation}
where $C_0$ is the correlation between nearby spins.  In this scenario, $N\langle C^2\rangle$ indeed is finite at large $N$, and the entropy is extensive, as it should be.

Because neurons connect through structures (axons and dendrites) that can be much longer than the spacing between neurons, many interesting biological networks or sub--networks do not have a clear notion of geometry or locality.  The result is that correlations need not have a systematic dependence on the distance between cells, and so the system is more nearly mean--field--like.  In a truly mean--field system, all $C_{\rm ij}$ would be drawn from the same distribution, and to enforce extensivity would require this distribution to have $\langle C^2 \rangle \propto 1/N$.  But, again, $\langle C^2 \rangle$ is an experimentally accessible quantity.

If we have a system of $N$ neurons with connections that reach across the entire network, it is plausible that recording from two neurons at random we will measure a correlation coefficient that is independent of the distance between cells and represents a sample from the overall distribution $P(C)$.  In the salamander retina, for example, there is no systematic relationship between correlations and distance as long as we stay within a radius of $\sim 200\,\mu{\rm m}$, and within such a correlated patch there are $N\sim 200$ cells \cite{puchalla+al_05}.   In such networks, $\langle C^2\rangle$ is a number we can measure by sampling many pairs of cells, even if we can never record from all $N$ cells simultaneously.

If we imagine networks with different values of $N$ but the same value of $\langle C^2\rangle$, corresponding to what we measure in a real network, this family of hypothetical networks will have an entropy per spin that varies with $N$, even at large $N$.  In this sense, there is no simple thermodynamic limit.  We can think about increasing $N$ at fixed $\langle C^2\rangle$ as being like changing temperature, as in the qualitative discussion of Ref \cite{schneidman+al_06}, or we can try to estimate the actual value of $N\langle C^2\rangle$ in the real system, and imagine a system in which $N\rightarrow \infty$ but $N\langle C^2\rangle$  is fixed to its experimental value.  A key point, which will be reinforced by the more detailed calculations below, is that $N\langle C^2\rangle$ can be large even when all $C_{\rm ij}$ are small, so that the impact of correlations on the entropy per spin depends on the size of the system.

\section{Results for a network of real neurons}

Interest in maximum entropy approaches to real biological networks was stimulated by Ref \cite{schneidman+al_06}, which analyzed the responses of neurons in a small patch of the salamander retina as they responded to a naturalistic movie.  The experiment used an array of electrodes to record from forty neurons within a radius of $\sim 200\,\mu{\rm m}$, a region throughout which there is no systematic dependence of correlations on distance, as described above.  It is reasonable, then, to view this experiment as a sample from the $\sim 200$ neurons in the patch.  In this sample, the distribution of correlation coefficients is peaked near zero, with almost all the weight at $C < 0.1$; a substantial fraction of correlation coefficients are negative, and the experiment is long enough that the threshold for statistical significance is  $|C|\sim 0.001$.  These weak pairwise correlations coexist with signatures of interesting collective behavior, such as a long tail in the probability that $K$ of the $N$ cells spike simultaneously, and dramatic discrepancies between the probability of different 10--cell patterns of response and the probabilities predicted if each cell were acting independently.  These discrepancies are resolved in the maximum entropy model.  This detailed analysis of patterns of response in small sub--networks was extended to groups of 20 and 40 cells \cite{tkacik+al_06,tkacik+al_09}, showing for example that three--point correlations are well predicted  from the maximum entropy model that incorporates only pairwise interactions.  These successes invite extrapolation to the behavior of larger groups of cells, where collective effects are predicted to be even more dramatic \cite{schneidman+al_06,tkacik+al_06,tkacik+al_09}.

\begin{figure}
\vspace{-2cm}
\includegraphics[width=\linewidth]{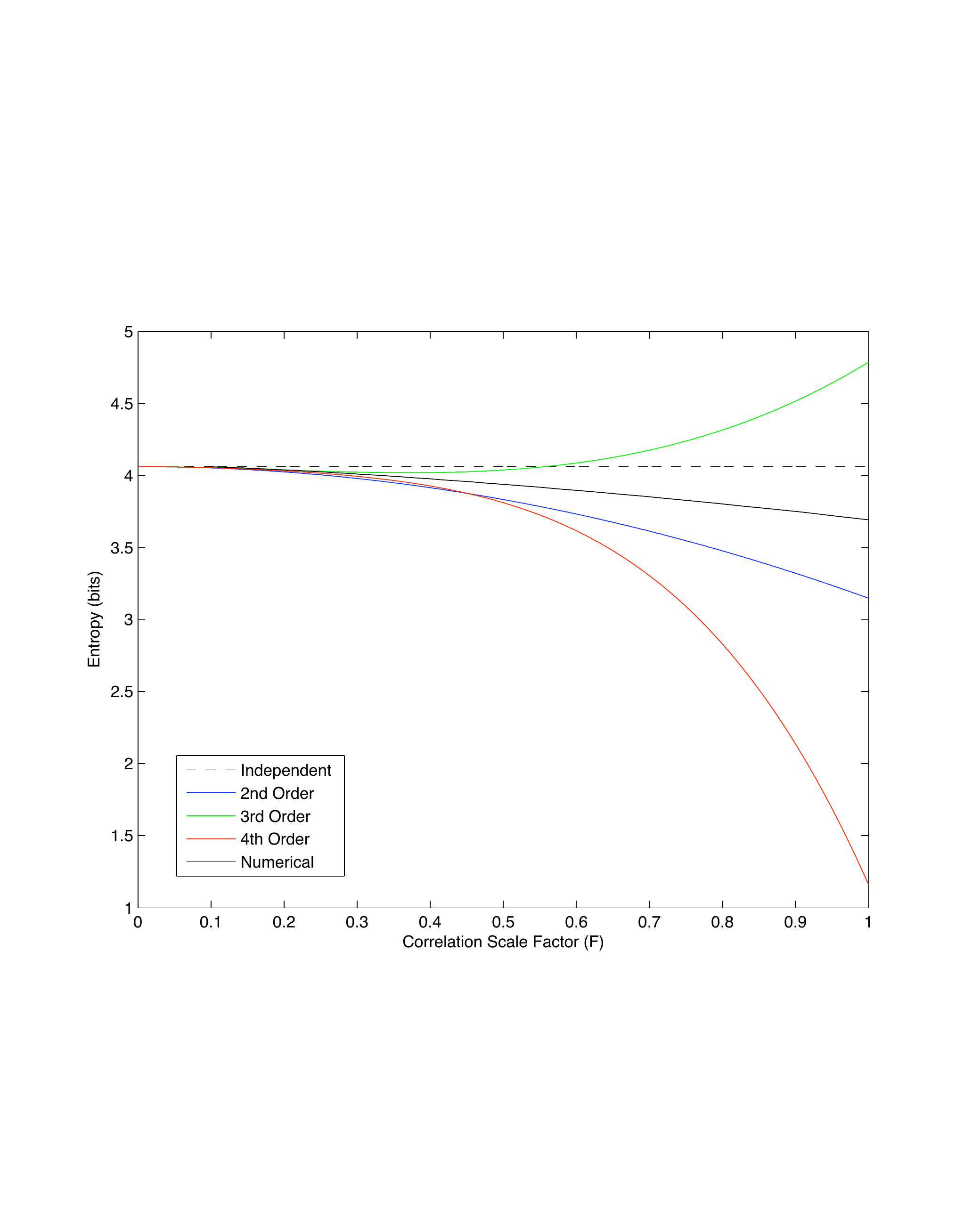}
\vspace{-3cm}
\caption{Entropy vs the strength of correlations for 20 cells.  As explained in the text, we consider a population of neurons with measured mean spike rates $\{\langle \sigma_{\rm i}\rangle\}$ and correlation coefficients  scaled by a factor $F$, $C_{\rm ij}\rightarrow FC_{\rm ij}$.  `2nd order' refers to the entropy to second order in perturbation theory i.e., $S(\{\langle\sigma_{\rm i}\rangle^{(0)}, FC_{\rm ij}\})=S_{0}(\{\langle\sigma_{\rm i}\rangle^{(0)}\})+\Delta S_{2}(\{FC_{\rm ij}\})$ and similarly for the other orders. We note that at $F\sim 0.5$, $\Delta S_{3}(\{\langle\sigma_{\rm i}\rangle^{(0)}, FC_{\rm ij}\})$ is roughly the same size as $\Delta S_{2}(\{FC_{\rm ij}\})$---perturbation theory is breaking down. Correlations for which $F\gtrsim 0.5$ can thus be considered to be large.   \label{EntropyvsF20CellsIndExact}}
\end{figure}

Here we are interested in reanalyzing the data of Ref \cite{schneidman+al_06} using our perturbation theory for the entropy.  We lean on the results of Ref  \cite{tkacik+al_06}, where numerical methods were used to construct the pairwise maximum entropy models for groups of $N=20$ neurons (exactly) and $N=40$ neurons (approximately, matching the measured $C_{\rm ij}$ within $\sim 1\%$).  These results give us essentially exact answers for the (maximum) entropy in these groups of cells, against which our perturbative results can be compared.

\begin{figure*}
\vspace{-2cm}
\includegraphics[width=0.32\linewidth]{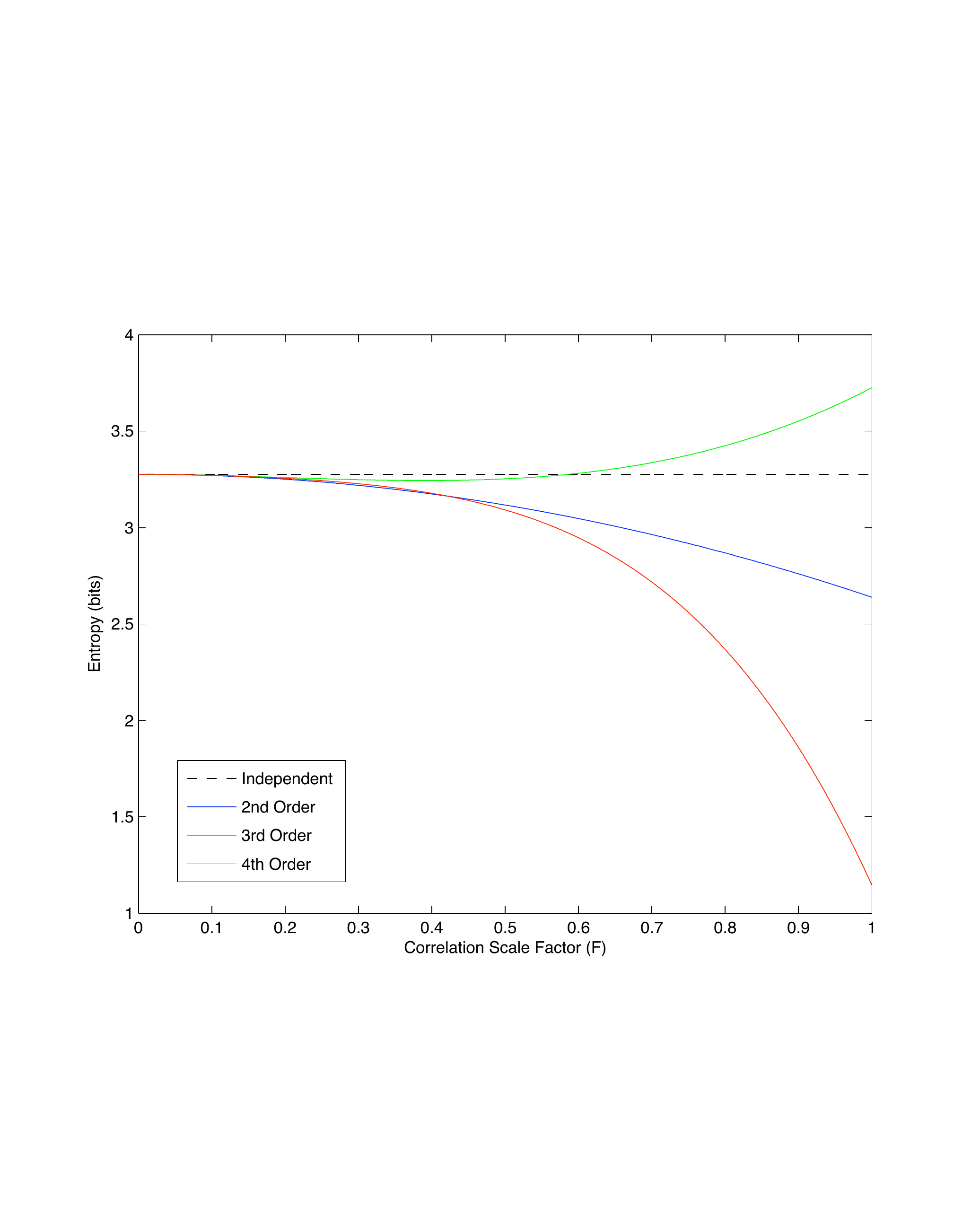}
\includegraphics[width=0.32\linewidth]{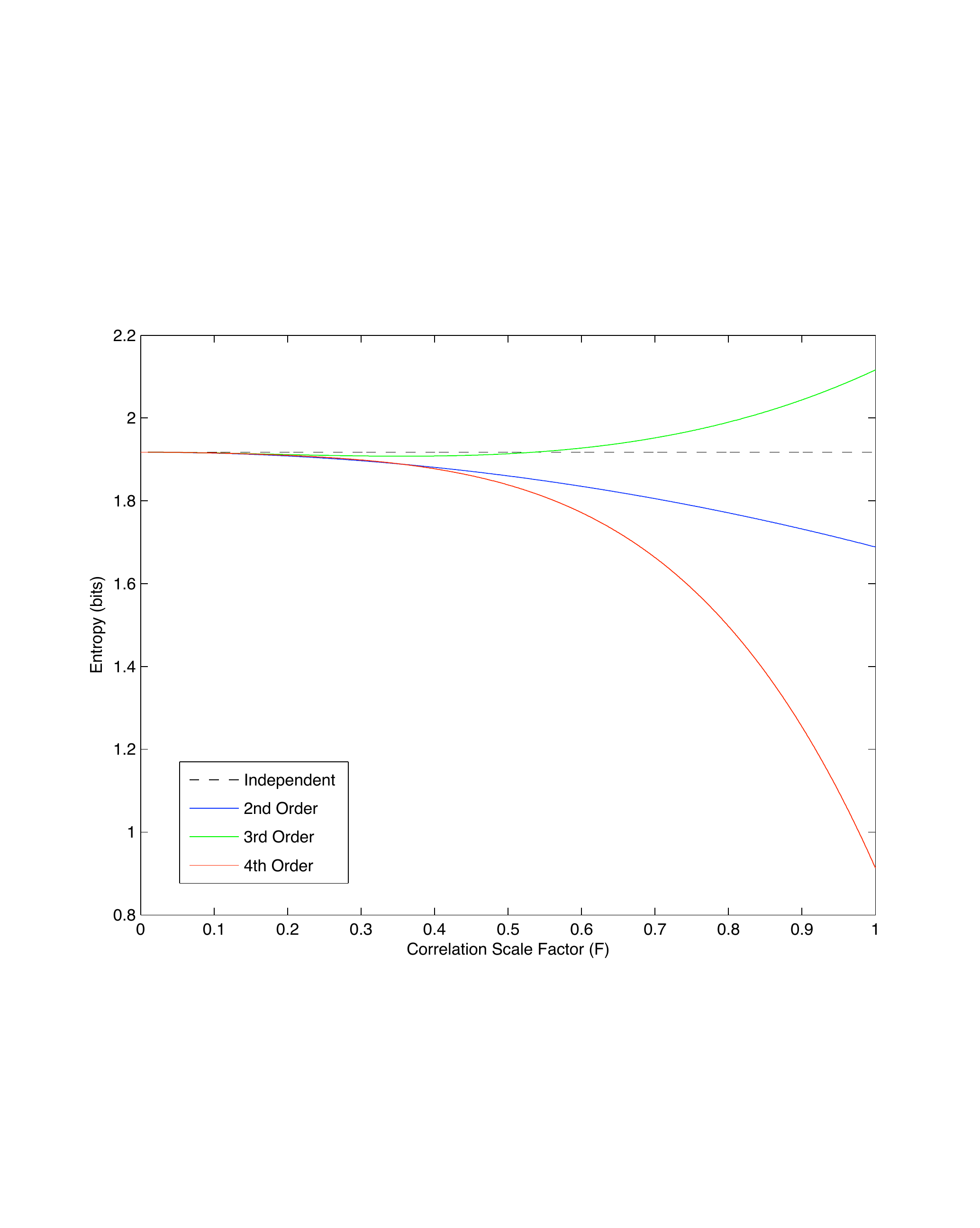}
\includegraphics[width=0.32\linewidth]{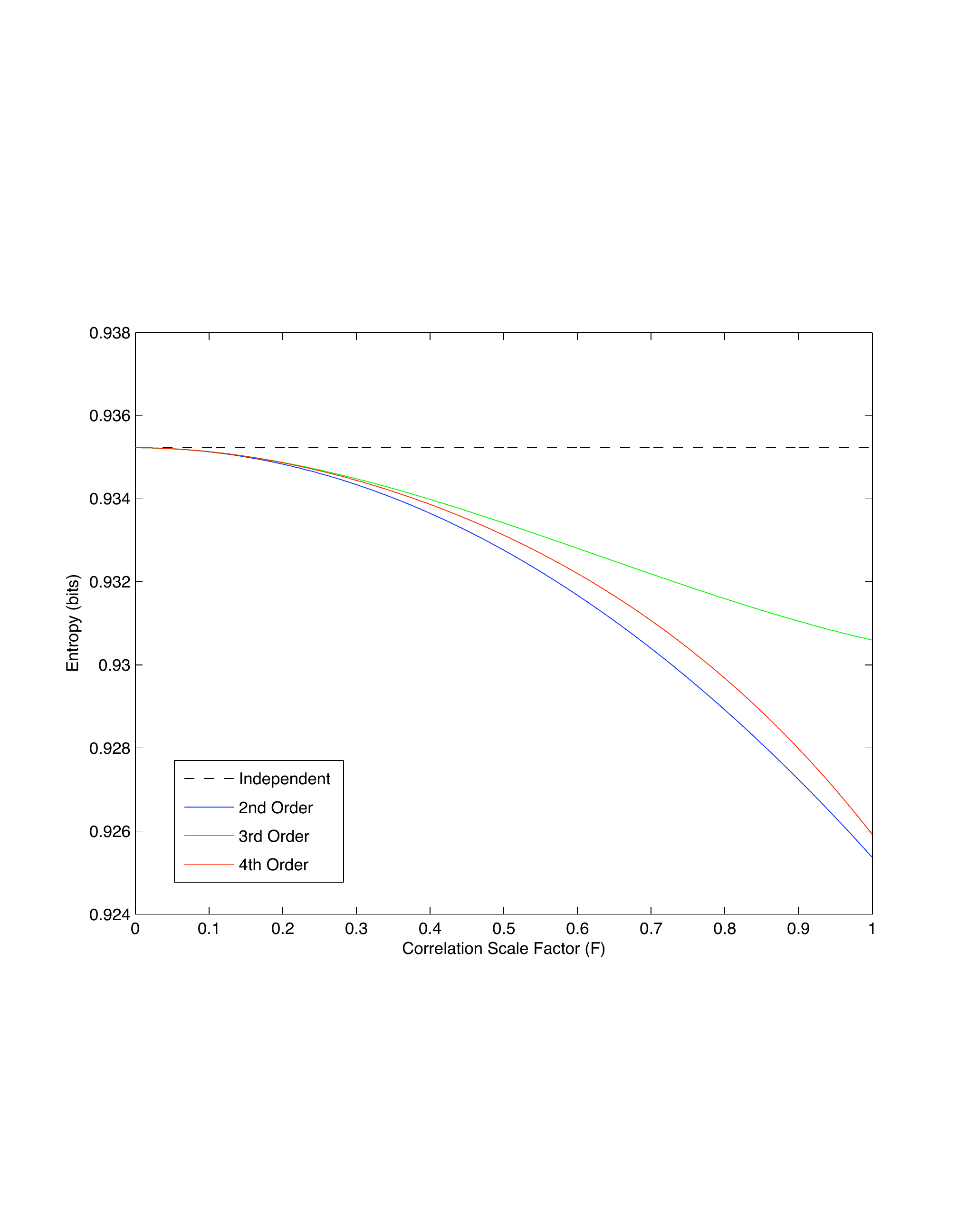}
\vspace{-1.25cm}
\caption{Entropy vs $F$ for (left to right) 15, 10 and 5 cells.
These groups are nested subsets of the 20 cells used in Figure \ref{EntropyvsF20CellsIndExact}.
\label{EntropyvsF15Cells}}
\end{figure*}

We would also like to have an internal standard for the validity of perturbation theory.  As usual, we can obtain such a criterion by asking whether successive orders of perturbation theory provide progressively smaller corrections, in this case to the entropy.  To gain control of the calculation, we imagine a population of neurons in which all correlation coefficients have been scaled by a factor $F$, $C_{\rm ij} \rightarrow F C_{\rm ij}$, but the mean spike rates (the expectation values $\langle \sigma_{\rm i}\rangle$) are held fixed.  Certainly as $F\rightarrow 0$ perturbation theory should work, and as we increase $F\rightarrow 1$ we approach the real system.  In Fig \ref{EntropyvsF20CellsIndExact} we present the entropy as a function of $F$ for a group of $N=20$ cells, as calculated in different orders of perturbation theory, comparing the exact results \cite{note2}.

We see from Fig \ref{EntropyvsF20CellsIndExact} that, at $F=1$, the third and fourth order contributions to  the entropy  overcorrect the second order approximation. The perturbative formalism at this scale of the correlations lies outside its range of validity. Scaling $F\to 0$, we see gross agreement between the perturbative results and the numerical result with convergence at roughly $F=0.3$. Comparing only successive contributions to the deviation of the entropy from the independent entropy we note that the magnitude of the third order correction $\vert\Delta S_{3}(\{\langle\sigma_{\rm i}\rangle^{(0)}, FC_{\rm ij}\})\vert$ is roughly the same as the magnitude of the second order correction $\vert\Delta S_{2}(\{FC_{\rm ij}\})\vert$ at $F\sim 0.5$. Qualitatively then, it is at these values of the correlations that the perturbative formalism for $N=20$ cells is breaking down and thus for $F\gtrsim 0.5$, the correlations are effectively strong.

As discussed in relation to the thermodynamic limit, our perturbation series mixes a dependence on the correlations themselves with a dependence on the size of the system.  If the scale of correlations is held fixed (for example, at the experimentally observed values!), then convergence of the series depends upon $N$.  At smaller $N$, we expect that the perturbative approach will work for larger values of the correlations.  To see this, we explicitly consider subsets of $15$, $10$, and $5$ cells out the $20$ we have analyzed so far.  For these different values of $N$ we again trace the perturbative predictions for the entropy as a function of $F$, which scales the correlations relative to their experimental values; results are shown in Fig \ref{EntropyvsF15Cells}.

For $15$ and $10$ cells, we do not see convergent behaviour of the series, and again the fourth order contribution $\Delta S_{4}(\{\langle\sigma_{\rm i}\rangle^{(0)}, FC_{\rm ij}\})$ significantly compensates for the third order contribution $\Delta S_{3}(\{\langle\sigma_{\rm i}\rangle^{(0)}, FC_{\rm ij}\})$. For $5$ cells, the series seems to be displaying convergent behaviour, with each successive correction representing some fraction of the last one. Also, in this case the series makes the sensible prediction that the entropy of the correlated system is smaller than when the correlations are zero; even this basic fact seems outside the reach of perturbation theory at larger $N$.

Although it might be interesting to know the particular answer for the entropy in specific groups of neurons, we are more interested in the overall validity of our perturbative approach.  As suggested above, we can think of any $N$ cells we study as being drawn out of a larger population, and in this population we can compute averages of the correlations in the combinations that enter the series for the entropy; our first example above was in Eq (\ref{avgC2}), and we can do this for every term in the series.  Up to third order, this yields
\begin{widetext}
\begin{eqnarray}\label{largeNseries}
S(N,F)&=& N\left(1+\frac{1}{\ln 2}\langle\ln\left(\cosh\left(\tanh ^{-1}(\langle\sigma_{\rm i}\rangle^{(0)})\right)\right)\rangle
-\frac{1}{\ln 2}\langle\langle\sigma_{\rm i}\rangle^{(0)}\tanh^{-1}(\langle\sigma_{\rm i}\rangle^{(0)})\rangle\right)\nonumber \\
 & &-\frac{1}{4\ln 2}N(N-1)F^{2}\langle {C_{\rm ij}}^2\rangle_{\rm i\neq j}\nonumber\\
 & &+\frac{1}{3\ln 2}N(N-1)F^{3}\Bigg{\langle}{C_{\rm ij}}^3\frac{\langle\sigma_{\rm i}\rangle^{(0)}}{\left(\delta\sigma_{\rm i}\right)^{(0)}_{\rm rms}}\frac{\langle\sigma_{\rm j}\rangle^{(0)}}{\left(\delta\sigma_{\rm j}\right)^{(0)}_{\rm rms}}\Bigg{\rangle}_{\rm i\neq j}\nonumber\\
 & &+\frac{1}{3!\ln 2}N(N-1)(N-2)F^{3}\langle C_{\rm ij}C_{\rm jl}C_{\rm li}\rangle_{\rm i\neq j, j\neq l, l\neq i}\;\;\textrm{bits}
\end{eqnarray}
\end{widetext}
where again, the averages $\langle\dots\rangle$ above are taken empirically.

\begin{figure*}
\vspace{-2cm}
\includegraphics[width=0.49\linewidth]{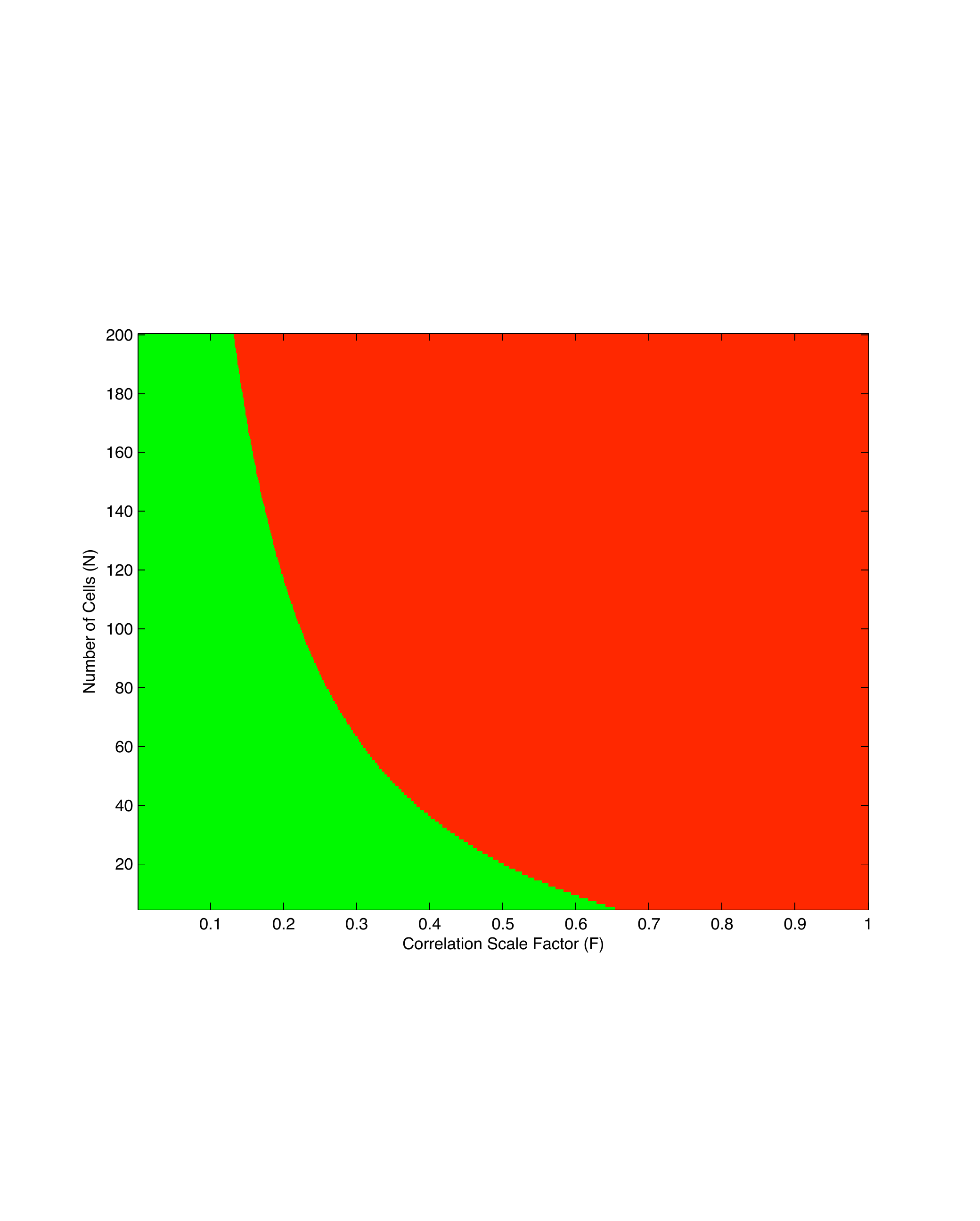}
\hfill
\includegraphics[width=0.49\linewidth]{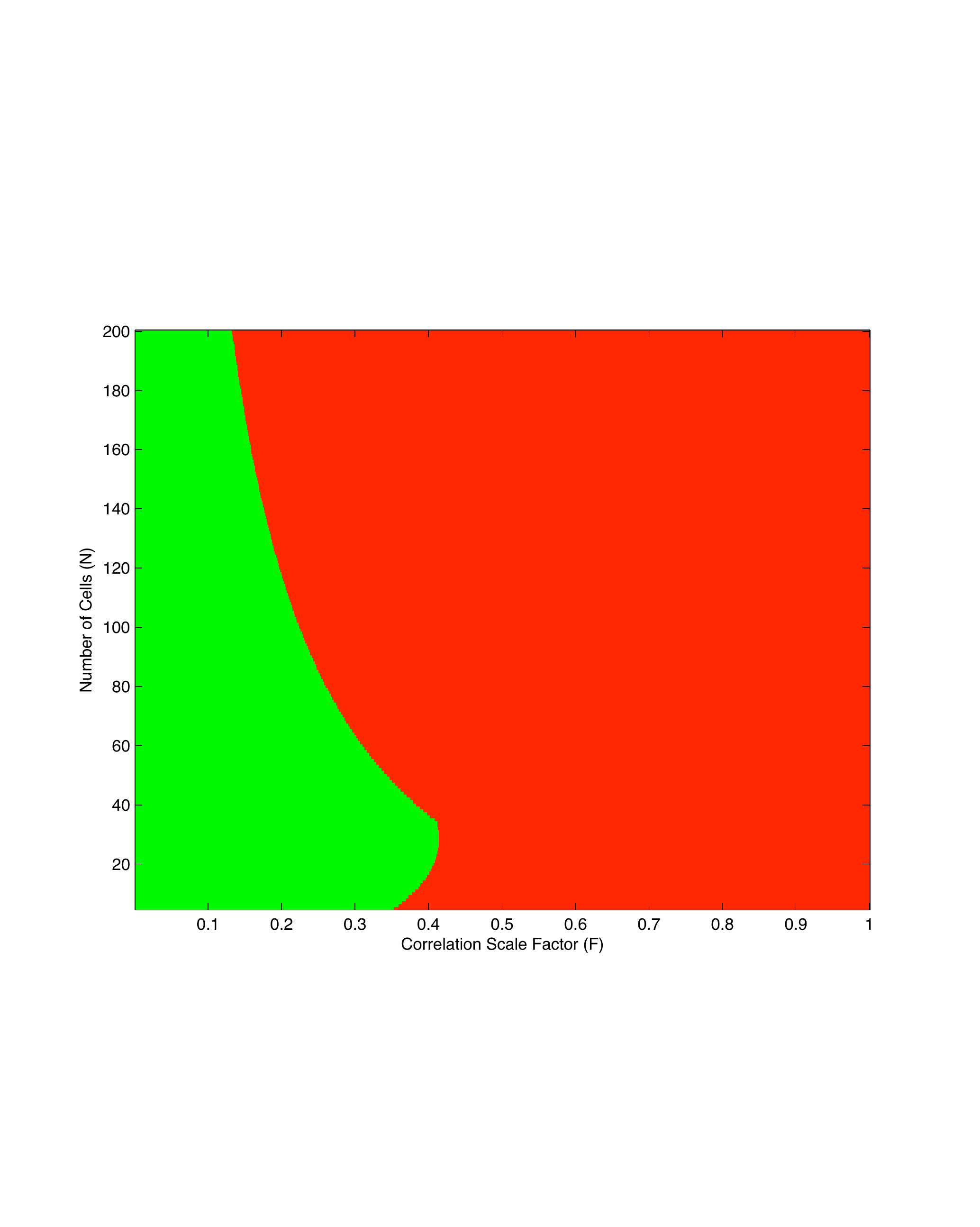}
\vspace{-2cm}
\caption{Regime of validity (green) for the perturbative entropy $S(N,F)$ expressed as a function of the number of cells $N$ and the correlation scale factor $F$.  At left, results to third order in perturbation theory. The green region corresponds to that part of our configuration space where the magnitude of the \emph{third} order correction is less than 90\% of the magnitude of the \emph{second} order correction $(|\Delta S_{3}(N,F)| < 0.9 |\Delta S_{2}(N,F)|)$ \emph{and} the total correlated entropy is less than the independent entropy. We have not included the fourth order term $\Delta S_{4}(N,F)$ in our considerations here. At right, green region corresponds to that part of our configuration space where the magnitude of the \emph{fourth} order correction is less than 90\% of the magnitude of the \emph{third} order correction which is in turn less than 90\% of the magnitude of the \emph{second} order correction ($|\Delta S_{4}(N,F)| < 0.9 |\Delta S_{3}(N,F)|\;,\;|\Delta S_{3}(N,F)| < 0.9 |\Delta S_{2}(N,F)|$) \emph{and} the total correlated entropy is less than the independent entropy.}
\label{regimeofvalidity}
\end{figure*}

Two conditions will guide us in constructing a sensible regime of validity for the series. The first is that, as with any perturbative series, successive corrections in the series must be less than some fraction of the previous order correction. We will use, for the sake of being concrete, a figure of 90\%. By this measure, the perturbative series will be said to have convergent behaviour at some combination of $N$ and $F$ if the magnitude of the $k$'th order correction $\vert\Delta S_{k}(N,F)\vert$ is less than 90\% of the magnitude of the $(k-1)$'th order correction $\vert\Delta S_{k-1}(N,F)\vert$ for all $k\geq 3$ included in the construction of the space.  We also insist that a valid perturbation theory must predict the entropy of the correlated system to be smaller than that of the uncorrelated ($F=0$) system, order by order.  With these criteria, we outline the regions of validity for perturbation theory in Fig \ref{regimeofvalidity}.  We emphasize that these results are a combination of theory with the empirical values for different moments of the correlation coefficients in the network of retinal neurons \cite{schneidman+al_06}.

\section{Discussion}

Most of what has been learned about the function of the brain, as well as other biological networks, has been learned by studying the activity of individual elements---the spikes generated by single neurons, the expression levels of single genes, the concentrations of particular metabolites, and so on.   Our intuition from statistical mechanics is that these large networks should have interesting collective behaviors.  A first step in searching for collective effects is to look for correlations between elements, and this has been explored in a wide variety of experiments; in the case of neural networks, this effort dates back roughly  forty years \cite{gerstein+clark_64,perkel+al_67b}.

It commonly is observed that correlations among neurons are weak but widespread.  Thus, almost all pairs of cells that plausibly are involved in the same neural functions have statistically significant, but small, correlations; examples include the retinal neurons considered here \cite{schneidman+al_06}, as well as in cerebral cortex \cite{cort-corr}.  If we ask about the implications of these weak correlations for the function of pairs of neurons, the answer must be that the effects are proportionally small.  But because the correlations are widespread, it is possible that the $\sim N^2$ correlated pairs add up to provide a signature of a qualitatively important collective effect.  Recent work has made this idea explicit, using the maximum entropy method to map data on the pattern of pairwise correlations into a statistical mechanics model of the whole network \cite{schneidman+al_06,tkacik+al_06,tkacik+al_09}.

Although the maximum entropy approach to describing networks of neurons has had some success, it could be that these successes are not probing a regime in which collective effects are possible.  In particular, if correlations are weak enough, one can imagine that a minimalist (i.e., maximum entropy) account of their impact succeeds, but for the trivial reason that all effects are minimal; this pessimistic claim has been made explicitly \cite{roudi+al_08}.  In this setting, pessimism about what we are learning from maximum entropy analyses of real neural data is equivalent to optimism about the utility of perturbation theory.  Perhaps amusingly, optimism about the detectability of collective behavior favored by physics--style models is equivalent to pessimism about the utility of one the physicists' favorite tools, perturbation theory.

Our main technical result is the development of a perturbation series that relates the maximum entropy to the observed pattern of pairwise correlations. To use this result, we imagine a population of $N$ neurons in which the distribution of mean spike rates is what one observes experimentally, and the distribution of correlations is as observed but scaled uniformly by a factor $F$.  Then we can study the entropy as a function of $N$ and $F$.  We note that the real system corresponds to $F=1$, and maximum entropy analyses have been pushed to $N=40$ using real data \cite{tkacik+al_06,tkacik+al_09}.  Figure \ref{regimeofvalidity} shows, unambiguously, that this is outside the regime in which we can expect the low orders of perturbation theory to give reliable answers.  Conversely, this means that the successes of the maximum entropy approach provide hints of interesting collective behavior, which is consistent with the observation of multiple locally stable states and an incipient critical point \cite{tkacik+al_06,tkacik+al_09}; for more on criticality in biological networks, see Ref \cite{mora+bialek_10}.

More generally, as we look at larger networks---perhaps the $\sim 10^2$ of transcription factors controlling gene expression in a single celled organism, or the tens of thousands of cells in small patch of visual cortex---the maximum correlations that can be captured in low orders of perturbation theory become smaller and smaller.  For the small patch of retina we have been discussing, the relevant $N\sim 200$, where the validity of perturbation theory is limited to $F\lesssim 0.1$, corresponding to correlations ten times smaller than what is seen experimentally.

Much of the work on correlations in biological networks is focused on the more limited question of whether a particular element in the correlation matrix is statistically significant.  Roughly speaking, if we make $K$ independent measurements, we expect that the threshold for statistical significance scales as $C^* \propto 1/\sqrt K$.  On the other hand, depending on  the pattern of correlations, the threshold for breakdown of perturbation theory can scale as $C_s \propto N$ or $C_s \propto \sqrt N$. If we are in the limit where $C_s < C^*$, we have a serious problem:  even `insignificant' correlations could be so large that they lead to a break down of perturbation theory.  Put another way, in this limit the statistical power of the experiments is so poor that it can fail to detect even the signatures of collective behavior in the network, let alone more subtle patterns of truly weak correlation.  This means that the number of measurements we need to make to provide a meaningful characterization of the correlations between two elements grows with the size of the network in which these elements are embedded.  Note that this is in contrast to the usual  statistical intuition, and provides a sobering message for experimentalists.

To summarize, the real patterns of pairwise correlations in biological networks---certainly the network of neurons that we consider in detail---fall outside the regime in which  the low orders of perturbation theory can capture their impact on the states of the network as a whole.  This is bad news for actually solving the inverse problem, but good news in that it means the successes of maximum entropy approaches to these networks are not simply the result of correlations being weak.

\begin{acknowledgments}
We thank MJ Berry II, E Schneidman, and G Tka\v{c}ik for helpful discussions.  Work in Princeton was supported in part by National Science Foundation Grants IIS--0613435 and PHY--0650617, and by the Swartz Foundation.  WB thanks his colleagues at the Kavli Institute for Theoretical Physics for their hospitality during the initial part of this work, and FA thanks the biophysics group at Princeton for providing a stimulating environment in which to continue the project.
\end{acknowledgments}

\begin{widetext}
\appendix
\section{Details of the expansion}

The pairwise maximum entropy partition function is given by Eq (\ref{isingdelta}), which we restate here for reference:
\begin{equation}\label{appisingdelta}
Z(\{h_{\rm i}, J_{\rm ij}\})=Z_{0}(\{h_{\rm i}^{0}\})\exp \left(-\sum_{\rm i}h_{\rm i}^{0}\langle\sigma_{\rm i}\rangle ^{(0)}\right)\Bigg\langle\exp\left(\sum_{\rm i}\delta h_{\rm i}\delta\sigma_{\rm i}+\sum_{\rm i < j}J_{\rm ij}\delta\sigma_{\rm i}\delta\sigma_{\rm j}\right)\Bigg\rangle ^{(0)}.
\end{equation}
We make the following identifications to align the form of our partition function [Eq (\ref{appisingdelta})] with that of the generic case discussed earlier:
\begin{displaymath}
g_{\alpha} = g_{\alpha}^{0}+\delta g_{\alpha}\equiv \left\{\begin{array}{ccc}
h_{\rm i}^{0}+\delta h_{\rm i} & \alpha \to {\rm i} & \sum_{\alpha}\to\sum_{\rm i}\\
0 + J_{\rm ij} & \alpha \to {\rm ij} &  \sum_{\alpha}\to\sum_{\rm i<j}
\end{array}\right.
\end{displaymath}
\begin{displaymath}
\Delta\hat{O}_{\alpha}\equiv \left\{ \begin{array}{cc}
\delta\sigma_{\rm i} & \alpha \to {\rm i}\\
\delta\sigma_{\rm i}\delta\sigma_{\rm j} & \alpha \to {\rm  ij}\;\;({\rm i < j})
\end{array}\right.
\end{displaymath}
We obtain an expression which deviates slightly from that of the corresponding equation for the generic case in Eq (\ref{partitionbeforecumulant}),
\begin{equation}
\label{appendixpart}
Z(\{g_{\mu}\})=Z_{0}(\{g_{\mu}^{0}\})\exp \left(-\sum_{\rm i}h_{\rm i}^{0}\langle\sigma_{\rm i}\rangle ^{(0)}\right)
\Bigg\langle\exp\left(\sum_{\alpha}\delta g_{\alpha}\Delta\hat{O}_{\alpha}\right)\Bigg\rangle ^{(0)}.
\end{equation}
Utilizing the definition of the cumulant expansion and noting that here, $\langle\Delta\hat{O}_{\alpha}\rangle ^{(0)}_{c}=\langle\Delta\hat{O}_{\alpha}\rangle ^{(0)}=0$ for all $\alpha$, we find that 
\begin{eqnarray}\label{cumulantisingpartition}
Z(\{g_{\mu}\})&=&Z_{0}(\{g_{\mu}^{0}\})\exp\left(-\sum_{\rm i}h_{\rm i}^{0}\langle\sigma_{\rm i}\rangle ^{(0)}\right)
\nonumber\\&&\,\,\,\,\,\,\,\,\,\times
\exp\Bigg{(}\frac{1}{2!}\sum_{\mu,\nu}\delta g_{\mu}\delta g_{\nu}\langle\Delta\hat{O}_{\mu}\Delta\hat{O}_{\nu}\rangle_{c}^{(0)}
+\frac{1}{3!}\sum_{\mu,\nu,\lambda}\delta g_{\mu}\delta g_{\nu}\delta g_{\lambda}\;\langle\Delta\hat{O}_{\mu}\Delta\hat{O}_{\nu}\Delta\hat{O}_{\lambda}\rangle_{c}^{(0)}+\cdots\Bigg{)}
\end{eqnarray}
where again $\langle\dots\rangle_{c}^{(0)}$ represents the cumulant of the enclosed operator with respect to the zeroth order distribution. One can show that equations Eq (\ref{yoyo}) to (\ref{finaldeltag}) hold now for the pairwise maximum entropy distribution as well. However in this instance we can elaborate on the final generic form for the perturbed couplings Eq (\ref{finaldeltag}), as one can explicitly calculate the susceptibility [Eq (\ref{susceptibility})] as follows,
\begin{equation}
\chi_{\mu\nu} \equiv \langle\Delta\hat{O}_{\mu}\Delta\hat{O}_{\nu}\rangle_{c}^{(0)}
= \langle\Delta\hat{O}_{\mu}\Delta\hat{O}_{\nu}\rangle ^{(0)}-\langle\Delta\hat{O}_{\mu}\rangle ^{(0)}\langle\Delta\hat{O}_{\nu}\rangle ^{(0)}.
\end{equation}
The indices $\mu$ and $\nu$ can take on forms $\rm i$ or $\rm ij$ giving us three unique combinations of indices for the quantity
$\chi_{\mu\nu}$. A short calculation shows that only cases where the same form of index appears in $\chi_{\mu\nu}$ does one obtain a nonzero value. Therefore one has that
\begin{equation}\label{ids}
\chi_{\mu\nu}=\delta_{\mu\nu}f(\mu)
\end{equation}
with no sum over the repeated index [on the right hand side of Eq (\ref{ids})], where $\delta_{\mu\nu}$ is the Kronecker delta function and
\begin{displaymath}
f(\mu) \equiv \left\{ \begin{array}{cc}
 \langle\left(\delta\sigma_{\rm i}\right) ^{2}\rangle ^{(0)}& \textrm{for $\mu = {\rm i}$}\\
 \langle\left(\delta\sigma_{\rm i}\right) ^{2}\rangle ^{(0)}\langle\left(\delta\sigma_{\rm j}\right) ^{2}\rangle ^{(0)}& \textrm{for $\mu = {\rm ij}$.}\\
\end{array}\right.
\end{displaymath}
Given these results, the expression for the perturbed couplings in Eq (\ref{finaldeltag}) becomes
\begin{eqnarray}
\delta g_{\mu}&=&\frac{\langle\Delta\hat{O}_{\mu}\rangle}{f(\mu)} - \frac{1}{2}\frac{1}{f(\mu)}\frac{\langle\Delta\hat{O}_{\beta}\rangle}{f(\beta)}\frac{\langle\Delta\hat{O}_{\gamma}\rangle}{f(\gamma)} \langle\Delta\hat{O}_{\mu}\Delta\hat{O}_{\beta}\Delta\hat{O}_{\gamma}\rangle_{c}^{(0)}\nonumber \\
& &- \frac{1}{3!}\frac{1}{f(\mu)}\frac{\langle\Delta\hat{O}_{\beta}\rangle}{f(\beta)}\frac{\langle\Delta\hat{O}_{\gamma}\rangle}{f(\gamma)}\frac{\langle\Delta\hat{O}_{\delta}\rangle}{f(\delta)}\langle\Delta\hat{O}_{\mu}\Delta\hat{O}_{\beta}\Delta\hat{O}_{\gamma}\Delta\hat{O}_{\delta}\rangle_{c}^{(0)}\nonumber \\
& &+ \frac{1}{2}\frac{1}{f(\mu)}\frac{1}{f(\gamma)}\frac{\langle\Delta\hat{O}_{\beta}\rangle}{f(\beta)}\frac{\langle\Delta\hat{O}_{\sigma}\rangle}{f(\sigma)}\frac{\langle\Delta\hat{O}_{\tau}\rangle}{f(\tau)}
\langle\Delta\hat{O}_{\mu}\Delta\hat{O}_{\beta}\Delta\hat{O}_{\gamma}\rangle_{c}^{(0)}\langle\Delta\hat{O}_{\gamma}\Delta\hat{O}_{\sigma}\Delta\hat{O}_{\tau}\rangle_{c}^{(0)}+\cdots.
\label{finaldeltagising}
\end{eqnarray}
where we sum over repeated indices and do not sum over indices in the argument of our function $f( \cdots )$. 

We can reconstruct Equation (\ref{S-intg}) for the entropy  in terms of the couplings by starting with
\begin{eqnarray}
S(\{g_{\mu}\}) &=& \ln Z(\{g_{\mu}\}) - g_{\mu}\frac{\partial\ln Z(\{g_{\mu}\})}{\partial (\delta g_{\mu})}.
\end{eqnarray}
Taking derivatives with respect to the perturbed couplings one finds that 
\begin{equation}
\frac{\partial S}{\partial \left(\delta g_{\beta}\right)}=-g_{\alpha}\frac{\partial\langle\Delta\hat{O}_{\alpha}\rangle}{\partial \left(\delta g_{\beta}\right)}
\end{equation}
which implies a relation for the entropy as a function of the couplings analogous to Eq (\ref{entropycoupling}),
\begin{equation}
\frac{\partial S}{\partial \langle\Delta\hat{O}_{\alpha}\rangle}=-g_{\alpha}\left(\{\langle\Delta\hat{O}_{\mu}\rangle\}\right).
\end{equation}
Integrating this equation from the set of zeroth order expectation values (the uncorrelated network) up to the experimental values of the correlations, one finds that 
\begin{equation}\label{entropyprelim}
S = S_{0}-g_{\alpha}^{0}\langle\Delta\hat{O}_{\alpha}\rangle-\int_{\{0\}}^{\{\langle\Delta\hat{O}_{\alpha}\rangle\}}d \langle\Delta\hat{O}_{\alpha}\rangle \delta g_{\alpha}\left(\{\langle\Delta\hat{O}_{\mu}\rangle\}\right)
\end{equation}
where $S_{0}$ is the entropy computed from the zeroth order partition function; namely, the entropy for the noninteracting or independent system. Again, the values of the one--point correlations $\{\langle\sigma_{\rm i}\rangle\}$ which one gleans from experiment are those that result in the no interaction limit, that is for $i=1,\dots,N$,
\begin{equation}\label{onepointassumption}
\langle\sigma_{\rm i}\rangle = \langle\sigma_{\rm i}\rangle ^{(0)}=\tanh(h_{\rm i}^{0}).
\end{equation}
This assumption forces $\langle\Delta\hat{O}_{\alpha}\rangle=0$ for $\alpha = {\rm i}$ and notably fixes the $N$ zeroth order couplings in the problem. Combining Eq (\ref{entropyprelim}) for the entropy, and Eq (\ref{finaldeltagising})  for the perturbed couplings, one obtains 
\begin{equation}\label{pairwisemaxentropy}
S = S_{0}+\Delta S\nonumber
\end{equation}
where for an array of $N$ cells, the independent entropy $S_{0}$, is given by
\begin{equation}\label{pairwisemaxentropyindependent}
S_{0} = N\ln 2+\sum_{i=1}^{N}\ln\left(\cosh\left(\tanh ^{-1}(\langle\sigma_{\rm i}\rangle ^{(0)})\right)\right)
-\sum_{i=1}^{N}\langle\sigma_{\rm i}\rangle ^{(0)}\tanh ^{-1}(\langle\sigma_{\rm i}\rangle ^{(0)})
\end{equation}
and the deviation of the entropy from this independent result is given by
\begin{eqnarray}
\Delta S &=& -\frac{1}{2}\frac{\langle\Delta\hat{O}_{\alpha}\rangle\langle\Delta\hat{O}_{\alpha}\rangle}{f(\alpha)} \nonumber\\
& &+ \frac{1}{3.2!}\frac{\langle\Delta\hat{O}_{\alpha}\rangle}{f(\alpha)}\frac{\langle\Delta\hat{O}_{\beta}\rangle}{f(\beta)}\frac{\langle\Delta\hat{O}_{\gamma}\rangle}{f(\gamma)} \langle\Delta\hat{O}_{\alpha}\Delta\hat{O}_{\beta}\Delta\hat{O}_{\gamma}\rangle_{c}^{(0)}\nonumber \\
& &+ \frac{1}{4.3!}\frac{\langle\Delta\hat{O}_{\alpha}\rangle}{f(\alpha)}\frac{\langle\Delta\hat{O}_{\beta}\rangle}{f(\beta)}\frac{\langle\Delta\hat{O}_{\gamma}\rangle}{f(\gamma)}\frac{\langle\Delta\hat{O}_{\delta}\rangle}{f(\delta)}\langle\Delta\hat{O}_{\alpha}\Delta\hat{O}_{\beta}\Delta\hat{O}_{\gamma}\Delta\hat{O}_{\delta}\rangle_{c}^{(0)}\nonumber \\
& &- \frac{1}{4.2}\frac{1}{f(\lambda)}\frac{\langle\Delta\hat{O}_{\alpha}\rangle}{f(\alpha)}\frac{\langle\Delta\hat{O}_{\beta}\rangle}{f(\beta)}\frac{\langle\Delta\hat{O}_{\gamma}\rangle}{f(\gamma)}\frac{\langle\Delta\hat{O}_{\delta}\rangle}{f(\delta)}
\langle\Delta\hat{O}_{\alpha}\Delta\hat{O}_{\beta}\Delta\hat{O}_{\lambda}\rangle_{c}^{(0)}\langle\Delta\hat{O}_{\lambda}\Delta\hat{O}_{\gamma}\Delta\hat{O}_{\delta}\rangle_{c}^{(0)}+\cdots.
\label{pairwisemaxentropydeviation}
\end{eqnarray}
For notational simplicity we will rewrite Eq (\ref{pairwisemaxentropydeviation}), splitting it into its component parts as follows:
\begin{equation}
\nonumber\Delta S = \Delta S_{2}+\Delta S_{3}+\Delta S_{4}+\dots
\end{equation}
where
\begin{eqnarray}
\Delta S_{2}&\equiv&-\frac{1}{2}\frac{\langle\Delta\hat{O}_{\alpha}\rangle\langle\Delta\hat{O}_{\alpha}\rangle}{f(\alpha)}\\ 
\Delta S_{3}&\equiv&\frac{1}{3.2!}\frac{\langle\Delta\hat{O}_{\alpha}\rangle}{f(\alpha)}\frac{\langle\Delta\hat{O}_{\beta}\rangle}{f(\beta)}\frac{\langle\Delta\hat{O}_{\gamma}\rangle}{f(\gamma)} \langle\Delta\hat{O}_{\alpha}\Delta\hat{O}_{\beta}\Delta\hat{O}_{\gamma}\rangle_{c}^{(0)}\\
\Delta S_{4}&\equiv&\frac{1}{4.3!}\frac{\langle\Delta\hat{O}_{\alpha}\rangle}{f(\alpha)}\frac{\langle\Delta\hat{O}_{\beta}\rangle}{f(\beta)}\frac{\langle\Delta\hat{O}_{\gamma}\rangle}{f(\gamma)}\frac{\langle\Delta\hat{O}_{\delta}\rangle}{f(\delta)}\langle\Delta\hat{O}_{\alpha}\Delta\hat{O}_{\beta}\Delta\hat{O}_{\gamma}\Delta\hat{O}_{\delta}\rangle_{c}^{(0)}\nonumber \\
& &- \frac{1}{4.2}\frac{1}{f(\lambda)}\frac{\langle\Delta\hat{O}_{\alpha}\rangle}{f(\alpha)}\frac{\langle\Delta\hat{O}_{\beta}\rangle}{f(\beta)}\frac{\langle\Delta\hat{O}_{\gamma}\rangle}{f(\gamma)}\frac{\langle\Delta\hat{O}_{\delta}\rangle}{f(\delta)}
\langle\Delta\hat{O}_{\alpha}\Delta\hat{O}_{\beta}\Delta\hat{O}_{\lambda}\rangle_{c}^{(0)}\langle\Delta\hat{O}_{\lambda}\Delta\hat{O}_{\gamma}\Delta\hat{O}_{\delta}\rangle_{c}^{(0)}
\end{eqnarray}
are the second, third and fourth order contributions to the total deviation of the entropy from its independent value respectively.  

To complete the calculation, one can explicitly compute the contribution of each of the terms in Eq (\ref{pairwisemaxentropydeviation}) to the deviation of the entropy from its independent value in Eq (\ref{pairwisemaxentropyindependent}) . For the sake of completeness we will outline this computation for the contribution of the third order term $\Delta S_{3}$ to $\Delta S$ and then simply state the result up to fourth order. 

\section{The third order contribution}

Consider the third order term $\Delta S_{3}$ (in nats),
\begin{equation}\label{thirdorderdeviation}
\Delta S_{3}= \frac{1}{3.2!}\frac{\langle\Delta\hat{O}_{\alpha}\rangle}{f(\alpha)}\frac{\langle\Delta\hat{O}_{\beta}\rangle}{f(\beta)}\frac{\langle\Delta\hat{O}_{\gamma}\rangle}{f(\gamma)} \langle\Delta\hat{O}_{\alpha}\Delta\hat{O}_{\beta}\Delta\hat{O}_{\gamma}\rangle_{c}^{(0)}.
\end{equation}
By assumption, $\langle\Delta\hat{O}_{\alpha}\rangle = 0$ for $\alpha = \rm i$ so the only terms which contribute to the sum above are those for which $\alpha = \rm ij$. Note also that  
\begin{equation}
\langle\Delta\hat{O}_{\alpha}\Delta\hat{O}_{\beta}\Delta\hat{O}_{\gamma}\rangle_{c}^{(0)} = \langle\Delta\hat{O}_{\alpha}\Delta\hat{O}_{\beta}\Delta\hat{O}_{\gamma}\rangle ^{(0)} + \textrm{(terms that vanish)}\nonumber.
\end{equation}
In the only nonzero sector of the above sum, we have, restoring the sums explicitly in our expression,
\begin{eqnarray}
\Delta S_{3} &= &\frac{1}{3!}\frac{1}{2^{3}}\sum_{\rm i\neq j\; k\neq l\; m\neq n}\langle\delta\sigma_{\rm i}\delta\sigma_{\rm j}\rangle\langle\delta\sigma_{\rm k}\delta\sigma_{\rm l}\rangle \langle\delta\sigma_{\rm m}\delta\sigma_{\rm n}\rangle
\frac{\langle\delta\sigma_{\rm i}\delta\sigma_{\rm j}\delta\sigma_{\rm k}\delta\sigma_{\rm l}\delta\sigma_{\rm m}\delta\sigma_{\rm n}\rangle ^{(0)}}{\langle\left(\delta\sigma_{\rm i}\right)^{2}\rangle ^{(0)}\langle\left(\delta\sigma_{\rm j}\right)^{2}\rangle ^{(0)}\langle\left(\delta\sigma_{\rm k}\right)^{2}\rangle ^{(0)}\langle\left(\delta\sigma_{\rm l}\right)^{2}\rangle ^{(0)}\langle\left(\delta\sigma_{\rm m}\right)^{2}\rangle ^{(0)}\langle\left(\delta\sigma_{\rm n}\right)^{2}\rangle ^{(0)}}.
\nonumber\\ &&
\end{eqnarray}
The correlation function
\begin{displaymath}
\langle\delta\sigma_{\rm i}\delta\sigma_{\rm j}\delta\sigma_{\rm k}\delta\sigma_{\rm l}\delta\sigma_{\rm m}\delta\sigma_{\rm n}\rangle ^{(0)}
\end{displaymath}
vanishes if the individual deviations $\delta\sigma_{\rm i}$ are left unpaired in the sum in which they reside. This means that one has two distinct contributions to the sum, one where the deviations $\delta\sigma_{\rm i}$ are paired off (e.g., $\rm i=k$, $\rm j=m$, $\rm l=n$) and one where three indices are equal (e.g., $\rm i=k=m$ and $\rm j=l=n$). The symmetry properties of the sum under investigation dictate that there are eight identical contributions to $\Delta S_{3}$ from the former and four identical contributions to $\Delta S_{3}$ from the latter. Thus we obtain
\begin{eqnarray}
\Delta S_{3}&=&\frac{1}{3!}\sum_{\rm i\neq j\; j\neq l\; i\neq l}\frac{\langle\delta\sigma_{\rm i}\delta\sigma_{\rm j}\rangle\langle\delta\sigma_{\rm i}\delta\sigma_{\rm l}\rangle \langle\delta\sigma_{\rm j}\delta\sigma_{\rm l}\rangle}{{\langle\left(\delta\sigma_{\rm i}\right)^{2}\rangle ^{(0)}}{\langle\left(\delta\sigma_{\rm j}\right)^{2}\rangle ^{(0)}}{\langle\left(\delta\sigma_{\rm l}\right)^{2}\rangle ^{(0)}}}
+\frac{4}{2.3!}\sum_{i\neq j}\frac{\langle\delta\sigma_{\rm i}\delta\sigma_{\rm j}\rangle ^{3}}{{\langle\left(\delta\sigma_{\rm i}\right)^{2}\rangle ^{(0)}}^{{3/2}}{\langle\left(\delta\sigma_{\rm j}\right)^{2}\rangle ^{(0)}}^{3/2}}\left[\frac{\langle\sigma_{\rm i}\rangle ^{(0)}\langle\sigma_{\rm j}\rangle ^{(0)}}{\sqrt{\langle\left(\delta\sigma_{\rm i}\right)^{2}\rangle ^{(0)}}\sqrt{\langle\left(\delta\sigma_{\rm j}\right)^{2}\rangle ^{(0)}}}\right]\nonumber .
\end{eqnarray}
Writing this in terms of the two point correlations $\{C_{\rm ij}\}$ introduced in Eq (\ref{correlations}), we obtain the following expression for the third order contribution to the entropy
\begin{eqnarray}
\Delta S_{3}&= &\Delta S_{3}(\{\langle\sigma_{\rm i}\rangle^{0}, C_{\rm ij}\})\nonumber\\
&= & \frac{1}{3!}\sum_{\rm i\neq j\; j\neq l\; i\neq l}C_{\rm ij}C_{\rm jl}C_{\rm li}
+\frac{1}{3}\sum_{\rm i\neq j}{C_{\rm ij}}^3\left[\frac{\langle\sigma_{\rm i}\rangle ^{(0)}}{\left(\delta\sigma_{\rm i}\right)^{(0)}_{\rm rms}}\right]\left[\frac{\langle\sigma_{\rm j}\rangle ^{(0)}}{\left(\delta\sigma_{\rm j}\right)^{(0)}_{\rm rms}}\right]
\end{eqnarray}
where $\left(\delta\sigma_{\rm i}\right)^{(0)}_{\rm rms} \equiv \sqrt{\langle\left(\delta\sigma_{\rm j}\right)^{2}\rangle ^{(0)}}$.

\clearpage
\section{The fourth order contribution}

One can employ a similar technique to that of the last subsection in computing the remaining terms in the construction of the entropy. Here we will simply state the final result for the entropy to fourth order in the correlation coefficients $\{C_{\rm ij}\}$:
\begin{eqnarray}\label{fourthorderentropy}
S&=&S(\{\langle\sigma_{\rm i}\rangle ^{(0)},C_{\rm ij}\})\nonumber \\
&=& N + \frac{1}{\ln {2}}\sum_{i=1}^{N}\ln\left(\cosh\left(\tanh ^{-1}(\langle\sigma_{\rm i}\rangle ^{(0)})\right)\right)-\frac{1}{\ln 2}\sum_{\rm i=1}^{N}\langle\sigma_{\rm i}\rangle ^{(0)}\tanh ^{-1}(\langle\sigma_{\rm i}\rangle ^{(0)})\nonumber \\
& &-\frac{1}{4 \ln 2}\sum_{\rm i\neq j}{C_{\rm ij}}^{2}\nonumber \\
& &+ \frac{1}{3!\ln 2}\sum_{\rm i\neq j\; j\neq l\; i\neq l}C_{\rm ij}C_{\rm jl}C_{\rm li}+ \frac{1}{3\ln 2}\sum_{\rm i\neq j}{C_{\rm ij}}^3\left[\frac{\langle\sigma_{\rm i}\rangle ^{(0)}}{\left(\delta\sigma_{\rm i}\right)^{(0)}_{\rm rms}}\right]\left[\frac{\langle\sigma_{\rm j}\rangle ^{(0)}}{\left(\delta\sigma_{\rm j}\right)^{(0)}_{\rm rms}}\right]\nonumber\\
& & - \frac{1}{24 \ln 2}\sum_{\rm i\neq j}{C_{\rm ij}}^{4}\left[\frac{1+9{\langle\sigma_{\rm i}\rangle^{(0)}}^{2}-3{\langle\sigma_{\rm j}\rangle^{(0)}}^{2}+9{\langle\sigma_{\rm i}\rangle^{(0)}}^{2}{\langle\sigma_{\rm j}\rangle^{(0)}}^{2}}{{\left(\delta\sigma_{\rm i}\right)^{(0)}_{\rm rms}}^{2}{\left(\delta\sigma_{\rm j}\right)^{(0)}_{\rm rms}}^{2}}\right]\nonumber \\
& &-\frac{1}{4\ln 2}\sum_{\rm i\neq j\; i\neq n\; j\neq n}{C_{\rm ij}}^{2}{C_{\rm in}}^{2}\nonumber\\
& &-\frac{1}{8\ln 2}\sum_{\rm i\neq j\; j\neq n\; n\neq l\;l\neq i\; i\neq n\; j\neq l}C_{\rm ij}C_{\rm jn}C_{\rm nl}C_{\rm li} + O(C^{5})\;\;{\rm bits}.
\end{eqnarray} 
\end{widetext}


\begin{thebibliography}{99}
\bibitem{note}
Our approach is closely related to that of Ref \cite{sessak+monasson_08}.
\end{thebibliography}
\end{document}